\documentstyle[epsf]{mn}

\author[L.V.E. Koopmans et al.]  
  {L.V.E.~Koopmans$^1$, A.G.~de
  Bruyn$^{1,2}$, D.R.~Marlow$^3$, N.~Jackson$^3$, \newauthor R.D.~
  Blandford$^5$, I.W.A.~Browne$^3$, C.D.~Fassnacht$^5$, S.T.~
  Myers$^4$,\newauthor T.J.~Pearson$^5$, A.C.S.~Readhead$^5$, P.N.~
  Wilkinson$^3$, D.~Womble$^5$\\ $^1$Kapteyn Astronomical Institute,
  P.O. Box 800, 9700 AV Groningen, The Netherlands\\ $^2$NFRA, P.O.
  Box 2, 7990 AA Dwingeloo, The Netherlands\\ $^3$University of
  Manchester, NRAL Jodrell Bank, Macclesfield, Cheshire SK11 9DL,
  England\\ $^4$Department of Physics, University of Pennsylvania,
  Philadelphia, PA 19104, USA\\ $^5$California Institute of
  Technology, Pasadena, CA 91125, USA}
\date{Accepted ; Received }
\pubyear{1998} 
\title{A new radio double lens from CLASS: B1127+385}

\begin{document}
\maketitle

\begin{abstract}

  We present the discovery of a new gravitational lens system with two
  compact radio images separated by 0.701$\pm$0.001 arcsec. The lens
  system was discovered in the Cosmic Lens All Sky Survey (CLASS) as a flat
  spectrum radio source. Both radio components show structure in a
  VLBA 8.4 GHz radio image. No further extended structure is seen in
  either the VLA, MERLIN or VLBA images. {\it Hubble Space Telescope
    (HST)} WFPC2 images in F555W and F814W show two extended objects
  close to the radio components, which we identify as two lens
  galaxies. Their colours and mass-to-light ratios seem to favour
  two late-type spiral galaxies at relatively high redshifts ($z_{\rm
    d}\ga0.5$).  Faint emission is also detected at positions
  corresponding to the radio images.

  A two-lens mass model can explain the observed VLBA structure. The
  best fit model has a reduced $\chi^2$ of 1.1. The relative positions
  of the VLBA subcomponents are reproduced within 0.08 mas, the flux
  density ratios within 0.19. We also reproduce the position angle and
  separation of the two VLBA subcomponents in A and B within the
  observational errors, which we consider strong evidence for the
  validity of the lens model. Moreover, we find a surface density axis
  ratio of $0.74^{+0.10}_{-0.12}$ for the primary lens (G1),
  consistent with the surface brightness axis ratio of $0.69\pm0.15$.
  Also, the surface density position angle of
  $(4.9^{+28.2}_{-22.4})^\circ$ of G1 compares well with the
  $(-6\pm13)^\circ$ position angle of the surface brightness
  distribution. The errors indicate the 99 per cent confidence
  interval.

\end{abstract}


\section{Introduction}

In the last few years gravitational lensing has proved useful not only
in the determination of cosmological parameters, such as the Hubble
constant (e.g. Refsdal 1964, 1966) and the cosmological constant (e.g.
Kochanek 1996), but also in the study of the mass distribution in the
universe and the mass distribution of lensing galaxies. To obtain a
sample of gravitational lens systems, relatively unbiased compared to
optical lens surveys (Kochanek 1991), which suffer from seeing effects
and dust obscuration, two large radio surveys, the Jodrell-Bank VLA
Astrometric Survey (JVAS; Patnaik et al. 1992; King et al., in
preparation; Wilkinson et al., in preparation) and the Cosmic Lens All
Sky Survey (CLASS; Myers et al., in preparation), were set up.
Together these surveys targeted $\sim$12,000 flat spectrum radio
sources with flux densities larger than 25 and 200 mJy for CLASS and
JVAS, respectively.  All sources were observed with the Very Large
Array (VLA) in A-array at 8.4 GHz with 0.2 arsec resolution. Objects
that showed signs of multiple compact components, or structure that
could be due to lensing, were listed for further high resolution radio
observations with MERLIN.  Those objects still exhibiting compact
structure in the MERLIN image were subsequently observed with the VLBA
to confirm their identification as a lens system and sometimes HST to
observe the optical emission of the lens galaxy and lens images.

In the following sections we give a detailed description of B1127+385,
a newly discovered gravitational lens system. In Section 2 we describe
the radio observations.  In Sections 3 the optical HST observations
are presented. In Section 4 we present a lens model based on the image
positions and flux density ratios from the VLBA observations. In
Section 5 we summarise our results and conclusions.

\section{Radio observations}

\begin{figure*}
\begin{center}
  \leavevmode
\vbox{%
  \vspace{0.5cm}
\hbox{%
  \epsfysize=8.0cm \epsffile{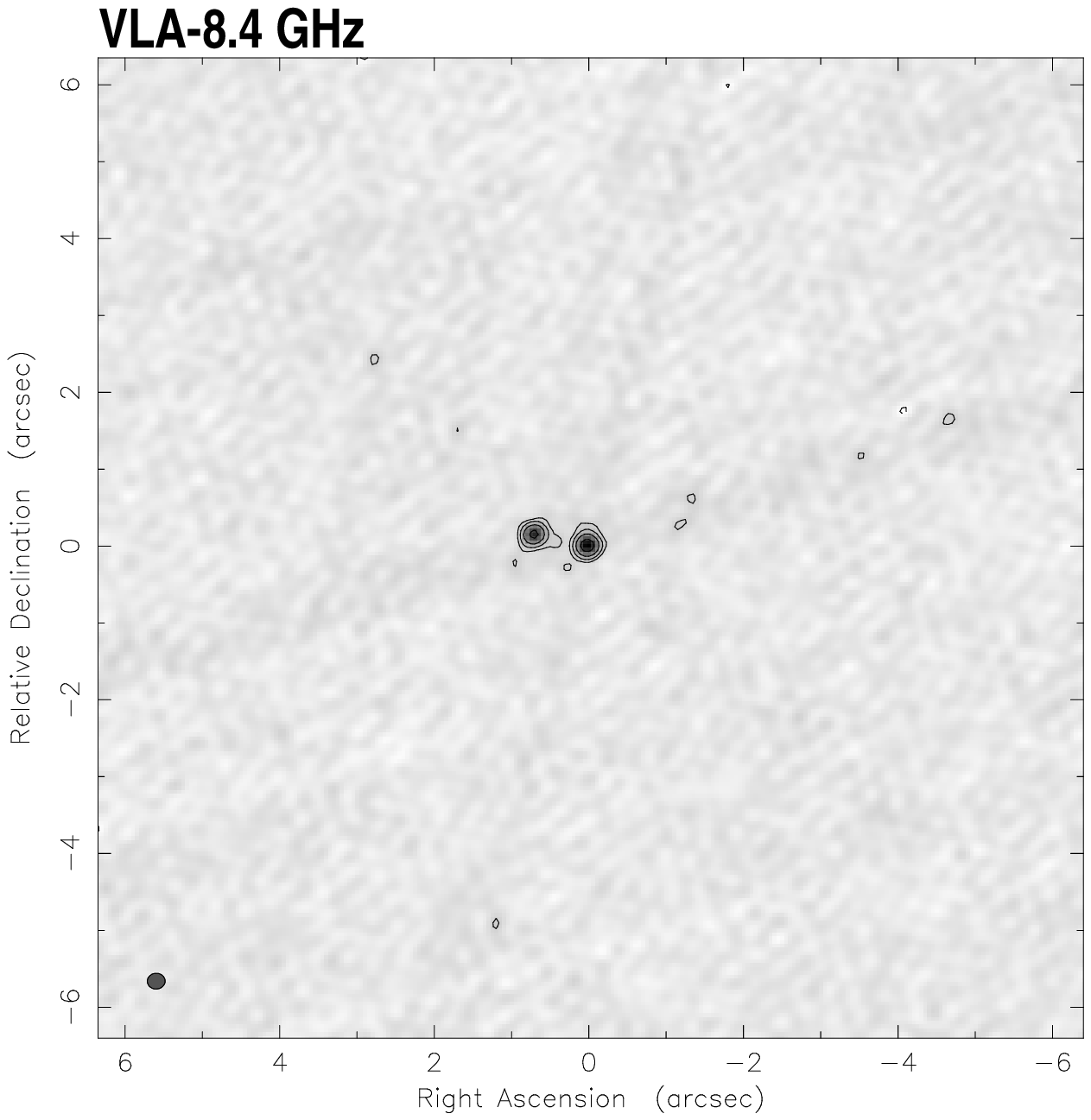} \hspace{1.5cm}
  \epsfysize=8.0cm \epsffile{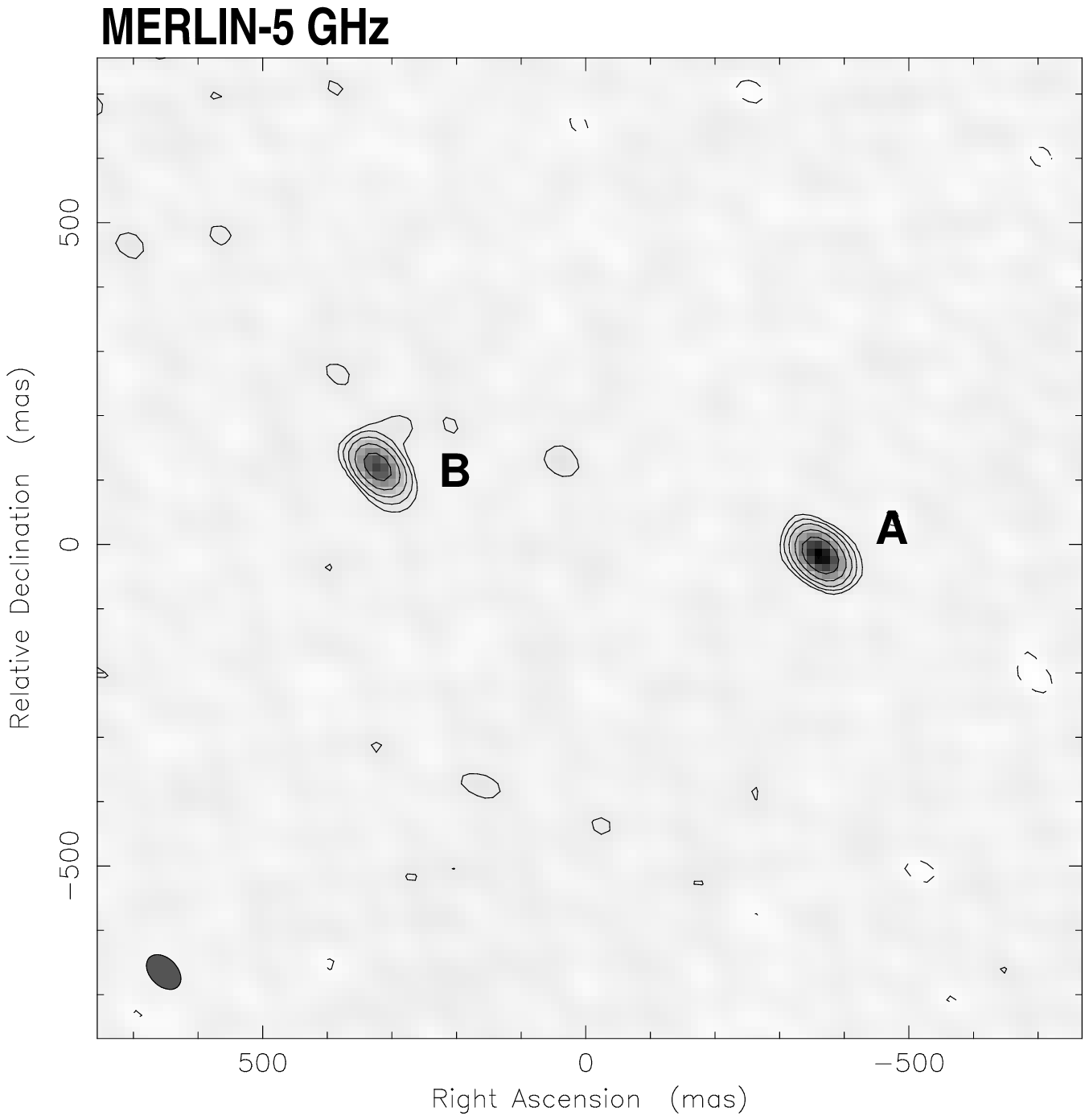}} \vspace{1cm}
\hbox{%
  \epsfysize=8.0cm \epsffile{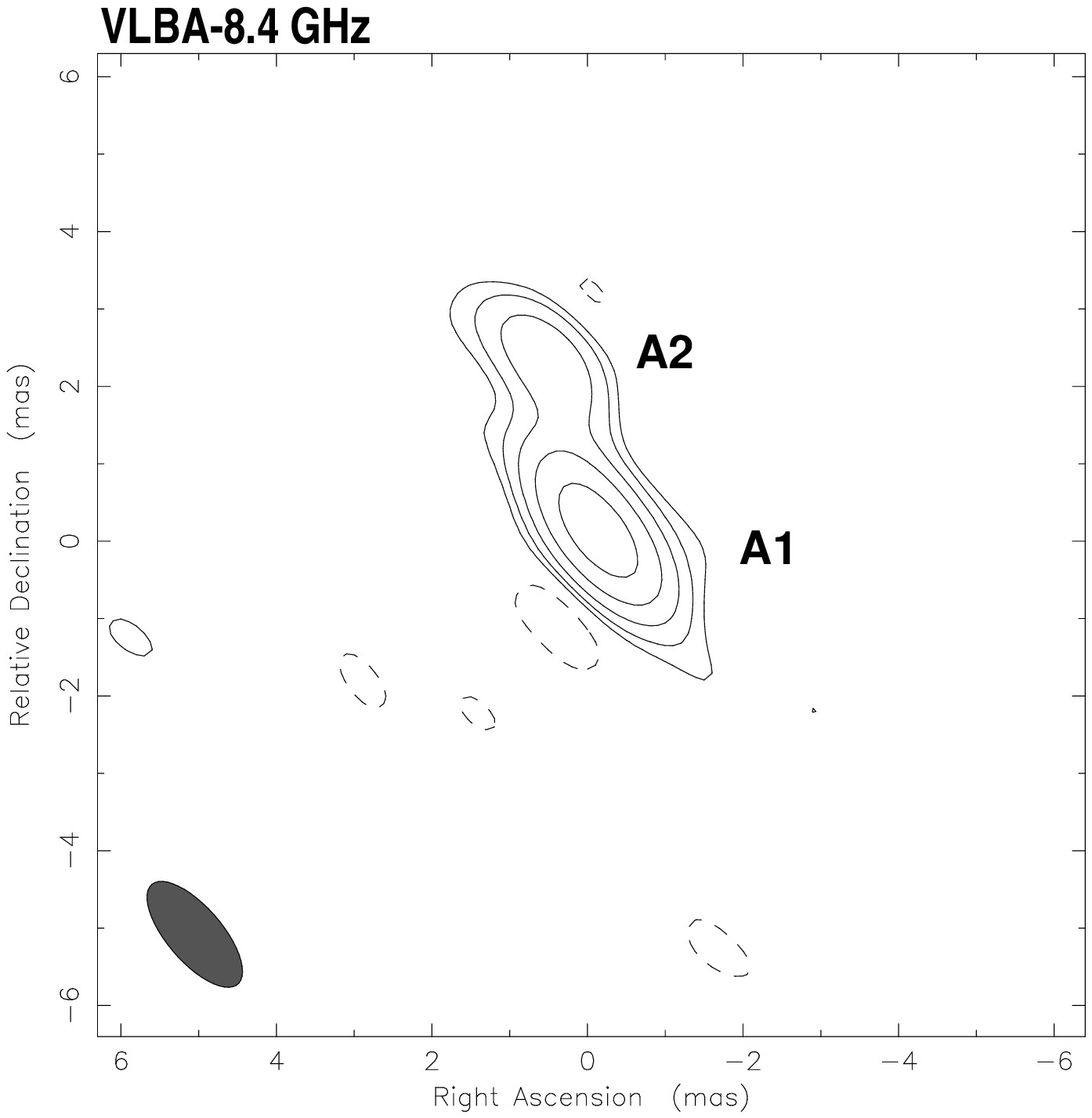} \hspace{1.5cm}
  \epsfysize=8.0cm \epsffile{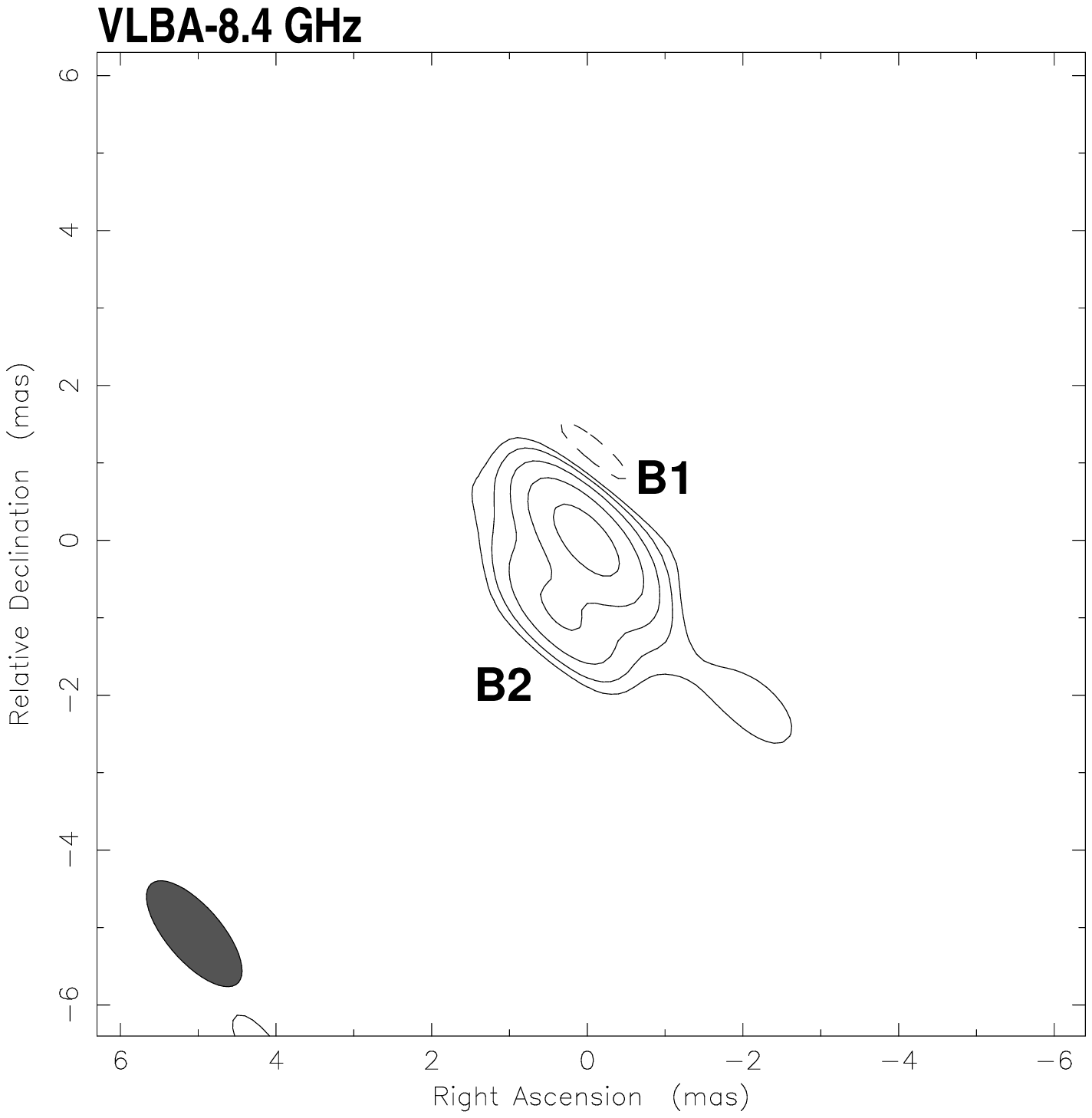}}}
\end{center}
\vspace{0.75cm}
\caption{Upper left: VLA 8.4 GHz snapshot of B1127+385 taken 1995
August 14. The image has a resolution of $0.2\times0.2$ arcsec and
shows two compact components (B is east, A is west) separated by 0.7
arsec.  Contours are at $(-3, 3, 6, 12, 24)\times 0.367$ mJy per beam.
Upper right: MERLIN 5 GHz observation taken 1996 December. The map has
a resolution of $0.06 \times 0.04$ arcsec (PA of 44$^\circ$). Contours
are at $(-3, 3,6,12,24,48,96) \times0.146$ mJy per beam. Lower left:
VLBA 8.4 GHz image of B1127+385 component A, taken 1996 November 4.
Contours are at $(-4,4,8,16,32,64)\times 0.090$ mJy per beam. Lower
right: Idem of component B. Contours are at
$(-4.5,4.5,9,18,36,72)\times 0.078$ mJy per beam. The VLBA maps have a
resolution of $1.7\times 0.7$ mas (PA of 41$^\circ$). }

\end{figure*}

\subsection{VLA and MERLIN observations}

B1127+385 was observed on 1995 August 14 with the VLA in A-array at
8.4 GHz as one of the $\sim$10,000 flat spectrum CLASS sources. The
image shows two compact components separated by $\sim$0.7 arcsec
(Fig. 1).  The 92-cm (0.327 MHz) WENSS (de Bruyn et al., in
preparation) flux density is $12\pm3$ mJy, establishing a slightly
inverted radio spectrum.  This implies that both components, with
similar flux density at 8.4 GHz, most likely have a flat or inverted
spectrum. This immediately made B1127+385 a strong lens candidate,
because a change alignment within 0.7 arcsec of two unrelated compact
flat spectrum radio sources is less than $10^{-6}$. Thus there is only
a probability $<$1 per cent of finding such a change alignment in the
sample of $\sim$10,000 flat spectrum sources.  No indication of
variability has been found sofar. A MERLIN 1.7 GHz long-track
observation on 1996 February 13 showed two compact components.
Subsequently, a snapshot observation with MERLIN was made at 5 GHz in
1996 December.  These observations show only the two compact
components A and B (Fig. 1). The VLA and MERLIN observations were
reduced in {\sc AIPS} and mapped by the Caltech package {\sc DIFMAP}
(Pearson et al. 1994; Shepherd 1997).  The flux densities and
positions were determined in {\sc DIFMAP} by fitting Gaussian models
to both components simultaneously. The results are listed in Table 1.
The spectral indices of the components are $-0.08$ (A) and 0.05 (B)
($S_\nu \propto \nu^{\alpha}$) respectively between the VLA 8.4 GHz
and MERLIN 5 GHz flux densities, and $-0.05$ (A) and $-0.11$ (B)
respectively between the VLA 8.4 GHz and MERLIN 1.7 GHz flux
densities. The errors on these spectral indices are $\sim0.15$.  The
compactness and similarity in spectral index of both radio components
underline the lens candidacy of B1127+385.  But high resolution radio
observations and optical follow-up are necessary to secure its lensing
nature.

\begin{table}
  \caption{VLA 8.4 GHz (1), MERLIN 5 GHz (2), MERLIN 1.7 GHz (3) and
    VLBA 8.4 GHz (4) astrometry and flux densities for B1127+385. The
    integrated flux densities from the VLBA observations are given.  The
    VLBA positions of A1 and B1 are $11^{\rm h}30^{\rm m}0.099^{\rm s},$
    $+38^\circ 12' 3.091''$ and $11^{\rm h}30^{\rm m}0.157^{\rm s},$
    $+38^\circ 12' 3.232''$ (J2000), respectively. The errors on the flux
    densities are $\sim$10 per cent. All errors are $1\sigma$.}
  \centering
  \begin{tabular}{lllcl}
    & $\Delta \alpha ({\rm mas})$ & $\Delta \delta ({\rm mas})$ &
    $S_{\nu} ({\rm mJy})$ & Instr.\\  A & $0\pm5$ & $0\pm5$ & 14.7
    & (1) \\ & $0\pm2$ & $0\pm2$ & 15.3 & (2) \\ & -- & -- & 16.0 & (3) \\ 
    A1 & $0.0\pm0.1$ & $0.0\pm0.1$ & 13.7 & (4) \\ A2 & $0.6\pm0.1$ &
    $2.1\pm0.1$ & $|$ &\\ B & $688\pm5$ & $145\pm5$ & 11.8 & (1)
    \\ & $685\pm2$ & $138\pm2$ & 11.5 & (2) \\ & -- & -- & 14.0 & (3)\\ B1 &
    $686.6\pm0.1$ & $140.1\pm0.1$ & 10.8 & (4) \\ B2 & $687.0\pm0.1$ &
    $139.2\pm0.1$ & $|$ & \\ 
  \end{tabular}
\end{table}

\subsection{VLBA observations}

VLBA 8.4 GHz observations were made on 1996 November 4. Phase
referencing was used, switching between B1127+385 (4 min integration)
and the strong nearby JVAS phase-reference source B1128+385 (Patnaik
et al.  1992; 2 min integration) for a period of 3.5 hours.  The map
of B1128+385 shows a sub-mas unresolved point source at 8.4 GHz.
Fringe-fitting (Thompson, Moran \& Swenson 1986) was therefore
performed on B1128+385 and the solutions were directly transferred to
B1127+385, which is $\sim$11 arcmin west of B1128+385. All data
reduction was performed in {\sc AIPS} and mapping was done in {\sc
DIFMAP}.  The data was uniformly weighted.  The resulting map
resolution is $1.7\times 0.7$ mas (PA of $41^\circ$).

Components A and B show clear evidence for the presence of
substructure (Fig. 1).  Model fitting in {\sc DIFMAP} shows that two
axisymmetric Gaussian components can well represent the substructure
in images A and B. The positions of the Gaussian components and their
flux densities are listed in Table 2 for their best fit (minimum
$\chi^2$).  The separation between A1 (A2) and B1 (B2) is 700.7
(700.0) mas and the position angle of the line from A1 to B1 is
$78.5^\circ$.  The flux density ratios between A1 and B1, and A2 and
B2 are 1.3 and 1.1, respectively. The integrated flux density ratio of
A over B is 1.3, consistent with the 8.4 GHz VLA (1.3) and 5 GHz
MERLIN (1.3) flux density ratios.  The large position angle difference
of $\sim 139^\circ$ between the subcomponents in A and B is expected
if A and B are lensed and given opposite parities (Schneider, Ehlers
\& Falco 1992).

\begin{table}
  \caption{The separations and position angles (north to east) of components 
    A2 (B2) with respect to components A1 (B1). The flux densities are
    those of (A1, A2) and (B1, B2), respectively. The errors on the
    flux densities are $\sim10$ per cent. All errors are $1\sigma$.}
  \centering
  \begin{tabular}{lccc}
    & Sep. (mas) & P.A. ($^\circ$) & $S_{\rm 8.4 GHz}$ (mJy) \\ 
    A & $2.20\pm0.15$ & $16\pm4$ & 10.5,3.2\\ B & $0.97\pm0.15$ &
    $155\pm9$ & 7.9,2.9\\ 
  \end{tabular}
\end{table}

\section{HST observations}

{\it Hubble Space Telescope (HST)} exposures of B1127+385 in the
filters F555W ($V$) and F814W ($I$) were taken on 1996 June 21, using
the Wide Field Planetary Camera (WFPC2). The exposures were taken on
the PC chip (45.5 mas pixel$^{-1}$) and the exposure times in $V$ and
$I$ band were $700{\rm s}+300$s and $2\times500$s, respectively.  A
standard reduction was performed on both images. The $I$-band image is
shown in Fig. 2. The $I$-band image shows two clear emission peaks
within $\sim$1 arcsec distance from the radio components.  Except for
a bright nearby galaxy 8 arcsec south, no other galaxies are seen near
B1127+385.  Because the absolute astrometry of the HST is poorly
matched (offsets of $\sim$1 arcsec) to the more accurate VLBI
astrometry, we cannot `blindly' overlay the optical and radio maps.
However, the contour plot of the optical $I$-band emission, convolved
to 0.1 arcsec (Fig. 3), clearly shows two bright (G1 and G2) and one
fainter emission feature. If we assume that the radio component A is
associated with the optical emission ($\sim9\sigma$ peak) west of G1,
we find that also radio component B is associated with an emission
feature ($\sim5\sigma$ peak). Although there appears to be a slight
offset between the optical emission and the radio position of B, this
could be due to the poorer signal-to-noise or the extended nature of
the optical emission near B.

\begin{figure}
\begin{center}
  \leavevmode
\vbox{%
  \epsfysize=6.5cm \epsffile{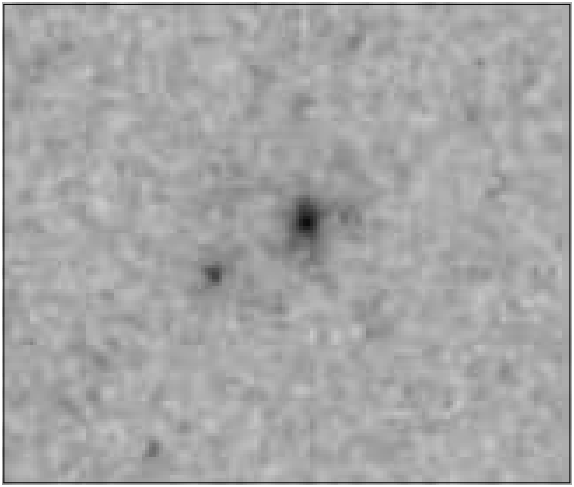}}
\end{center}
\caption{HST $I$-band image of B1127+385. The two radio components
  are associated with two emission features west (A) and east (B) of
  G1 (see also Fig. 3). North is up, east is left. The area shown is
  3.1 arcsec $\times$ 2.6 arcsec.}
\end{figure}

Both G1 and G2 are extended, suggesting that both are
galaxies. Photometry and relative astrometry were performed on G1 and
G2 in $I$ band, and only photometry in $V$ band (Table 3). The $V-I$
colour indices of G1 and G2 are 1.9 and 2.0 mag, respectively. The
separation is 0.60 arcsec and the position angle of the line G1-G2 is
$120^\circ$.

\begin{table}
  \caption{HST relative astrometry ($I$ band) and photometry for
    B1127+385.  Component G1 is located at $11^{\rm h}30^{\rm
      m}0.1726^{\rm s}$, $+38^\circ 12' 1.903''$ (according to the {\sc
      stsdas/metric} routine in {\sc IRAF}). All errors are $1\sigma$.}
  \centering
  \begin{tabular}{lrrcc}
    & $\Delta \alpha ({\rm mas})$ & $\Delta \delta ({\rm mas})$ &
    $I$ (magn.) & $V$ (magn.) \\ G1 & $0\pm5$ & $0\pm5$ &
    $22.5\pm0.1$ & $24.4\pm0.2$ \\ G2 & $521\pm5$ & $-299\pm5$ &
    $23.5\pm0.1$ & $25.5\pm0.2$ \\
  \end{tabular}
\end{table}

\begin{figure}
\begin{center}
  \leavevmode
\vbox{%
  \epsfysize=6.5cm \epsffile{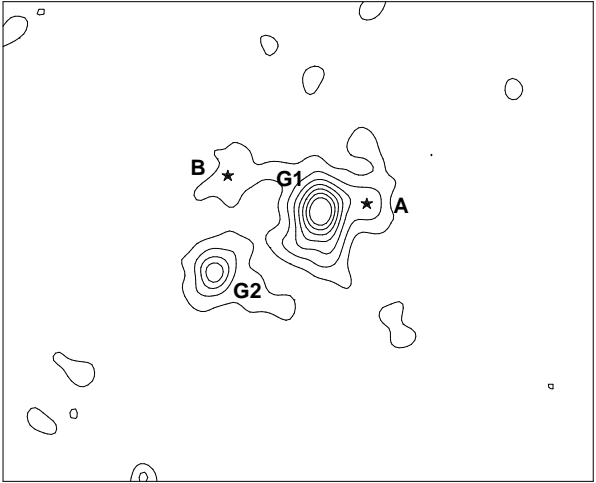}}
\end{center}
\caption{Contour plot of B1127+385.
  The HST image has been convolved to 0.1 arcsec to bring out the optical
  emission more clearly. The two markers show the radio positions of A
  and B, where A has been placed on the centre of the optical emission
  as determined by a Gaussian fitting procedure. The contours indicate
  $(3,6,9,12,15,18,21)\times{\rm rms}$ noise in the image.}
\end{figure}

%%%%%%%%%%%%%%%%%%%%%%%%%%%%%%%%%
%    Modelling of B1127+385     %
%%%%%%%%%%%%%%%%%%%%%%%%%%%%%%%%%

\section{Modelling}

In this section we present a model that reproduces the observed
properties of B1127+385. We use a Singular Isothermal Ellipsoid (SIE)
mass distribution (Kormann, Schneider \& Bartelmann 1994) to describe
the lens galaxies. We assume a smooth FRW universe. If not mentioned
otherwise, all errors indicate 99 per cent confidence intervals.

From the VLBA observations we obtain 10 constraints (8 from the image
positions and 2 from the flux density ratios). The two source
positions give 4 free parameters and the mass model gives 3 (velocity
dispersion, surface density (SD) axis ratio and position angle). The
number of degrees of freedom (NDF) is therefore 3.

\subsection{Single lens mass model}

Initially we try a model consisting of a single SIE galaxy. We place
the mass distribution on the surface brightness (SB) centre of G1,
determined by fitting a 2D-Gaussian profile to it. The position of G1
relative to the optical emission feature associated with radio
component A is $(-0.228'', -0.038'')$, with a $1\sigma$ error of
$\sim$5 mas in both $x$ and $y$. The parities of components A and B
are taken as --1 and +1, respectively. Choosing +1 and --1 for A and
B, and letting the centre of the SD distribution move freely, give in
all cases (both for the single and double lens case) unsatisfactory
models ($\chi^2\ga5$).

Using the image positions and flux density ratios of the radio
components from Tables 1 and 2, we project rays back on the source
plane through the lens. For the two image pairs -- A1\&B1 and A2\&B2
-- we simultaneously minimize the distance between the back-projected
rays and the difference between observed and model flux density ratios
(Kayser 1990). We allow for a $1\sigma$ error of 0.1 mas in the
relative $x$ and $y$ distances between the positions of the two
back-projected rays and their average position. A $1\sigma$ error of
0.15 is allowed for the flux density ratios. When the model has
converged sufficiently, we use the average source positions to
calculate the image positions in the lens plane.  These are
subsequently used to calculate a $\chi^2$ from the mismatch with the
observed image positions. The resulting mass model parameters for
minimum $\chi^2$ (lens plane) are listed in Table 4 (model I).

\subsection{Double lens mass model}

Although G1 is the primary lens, G2 cannot be neglected as it lies
close to G1 and is only $\sim$1 mag fainter.  We therefore extend the
model by placing a Singular Isothermal Sphere (SIS) at the position of
G2 --- $(-0.749'', -0.337'')$ relative to component A, with a
$1\sigma$ error of $\sim$5 mas in $x$ and $y$. We choose a SIS for G2,
because G2 is only an external perturber of the primary lens G1 and as
long as there is no need to complicate the model (poor $\chi^2$) one
should keep the mass model as simple as possible. Using a SIE for G2
would add two extra free parameter (axis ratio and position angle) and
make the mass model less constrained.

The velocity dispersion of G2 is fixed at $0.79\cdot
\sigma_\parallel^{\rm G1}$, using $L\propto\sigma_{\parallel}^4$
(e.g. Faber \& Jackson 1976) in combination with the $\sim$1 mag
difference between G1 and G2 and assuming they have similar
mass-to-light ratios. We repeat the $\chi^2$ minimization
procedure. The resulting model parameters are listed in Table 4 (model
II). Model II reproduces the relative positions of all four images to
within 0.08 mas, compared to 0.16 mas for model I. And although the
number of degrees of freedom does not change between models I and II,
the reduced $\chi^2$ of model II is significantly smaller than that of
model I. A reduced $\chi^2>4.5$ corresponds to a probability of
$4\cdot 10^{-3}$ and model I can therefore be rejected as an
appropriate model with 99.6 per cent confidence. So adding G2 improves
the mass model significantly. A faint image will be formed to the
south-east of G2, which can be removed by a very small core radius
(0.01$''$) for G2, without changing the model parameters at any
signifcant level. The critical and caustic structure of model II is
shown in Fig. 4, where G2 has been given a core radius of 0.01$''$.

If G1 and G2 are spiral galaxies, the relation between luminosity and
velocity dispersion is $L\propto\sigma_{\parallel}^{2.5}$ (Tully \&
Fisher 1977), hence $\sigma_\parallel^{\rm G2}\approx0.69\cdot
\sigma_\parallel^{\rm G1}$. The best model then has a slightly
increased $\chi^2$ of 1.6, still a considerable impovement over model
I. The resulting model parameters deviate by less than 3 per cent in
velocity dispersion and axis ratio from model II. The position angle
of G1 becomes $-5.3^\circ$, even in better agreement with the observed
surface brightness position angle (see below).  However, all values
are well within the errors determined by Monte-Carlo simulations for
model II (see Section 4.4 and Table 4).

Using a SIE for G2 would have the most influence on the position angle
and axis ratio of the mass model of G1. However, the close agreement
between these parameters and their observed values (see next
paragraph), as well as the small resulting minimum $\chi^2$, indicate
that adding extra free parameters is not necessary.

\begin{figure}
\begin{center}
  \leavevmode
\vbox{%
  \epsfysize=8.5cm \epsffile{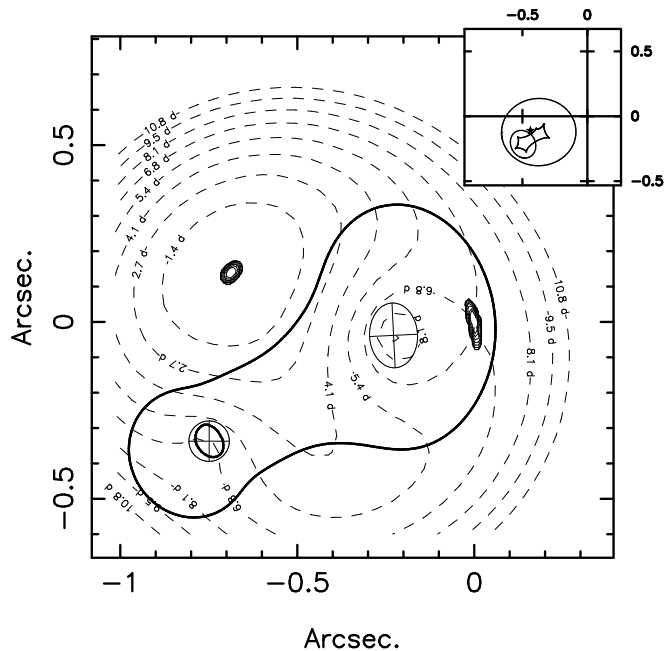}}
\end{center}
\caption{The critical curves (thick lines) and caustics (subpanel)
  of model II in Table 4. The star (subpanel) indicates the model
  position of the source as seen in the source plane. The solid contours
  depict the images at the positions of the radio components, if the
  source surface brightness is gaussian with a FWHM of 5 mas. The lens
  galaxy G1 is indicated by the cross-haired ellips, G2 by the
  cross-haired circle.  The dashed contours indicate the time-delay
  surface for $z_{\rm d}=0.5$, $z_{\rm s}=1.5$, H$_0$=50 km s$^{-1}$
  Mpc$^{-1}$, $\Omega_{\rm m}=1$ and $\Omega_{\Lambda}=0$.}
\end{figure}

\subsection{Surface density versus surface brightness}

The SD axis ratio of G1, $(b/a)_{\Sigma}^{\rm G1}= 0.74$ (model II),
is only slightly larger than the SB axis ratio of $0.69\pm0.15$ that
we find from fitting a 2D-Gaussian profile to the SB distribution of
G1. The same Gaussian fit gives a position angle of $(-6\pm13)^\circ$,
close to the SD position angle of $4.9^\circ$. Although the SB profile
of G1 is most likely not Gaussian, the position angle and axis ratio
inferred from a 2D-Gaussian fit will give a good indication of the
value for these parameters.

Strong evidence that B1127+385 is a gravitational lens system is given
by the expected centre of the SD distribution of G1. In Fig. 5 the 99
per cent confidence contour of the central SD position of G1 is
plotted. The two circles indicate the radio components A and B. If we
assume that the two faint optical emission features are associated
with A and B, we find that the optical emission peak of G1 (cross)
falls perfectly inside the 99 per cent confidence contour. In other
words, the position of the SB centre of G1 relative to the faint
emission feature west of it coincides with the position of the SD
centre of G1 relative to radio component A.  This suggests G1 and G2
are indeed two lens galaxies and the faint optical emission features
(Fig. 3) are associated with the radio components.

\subsection{Monte-Carlo simulations}

To investigate the overall reliability of model II, we performed Monte
Carlo simulations. We minimize $\chi^2$ for 10,000 models, where we
add Gaussian distributed errors ($1\sigma$) to the relative image
positions (0.1 mas), flux density ratios (0.15), galaxy positions (5
mas) and velocity dispersion ratio between G1 and G2 (0.10).

\subsubsection{Mass model parameters}

The parameter probability density distributions of all models with
$\chi^2<11.3$ (99 per cent confidence interval) are shown in
Fig. 6. The figure also shows the 99 per cent confidence interval
(shaded) of the observed SB axis ratio and position angle. We see that
the probability distributions of both the SD axis ratio and position
angle have considerable overlap with these shaded regions. Hence, the
axis ratio and position angle of the luminous matter agree well with
those inferred from the SD distribution of G1. The mean values of the
recovered position angle and separation of A1 relative to A2 are
$(18\pm7)^\circ$ and $2.0\pm0.2$ mas and for B1 relative to B2
$(163\pm3)^\circ$ and $1.1\pm0.2$ mas, where the errors are the rms
values of the parameter probability distributions. These recovered
model parameters compare well with the observed values listed in Table
2.  We consider this as strong evidence for the validity of the lens
model.

To investigate the ratio $(\sigma_{\parallel,\rm
G2}/\sigma_{\parallel,\rm G1})$, we calculate the minimum $\chi^2$ for
a range of this ratio, as shown in Fig. 6. The two horizontal lines
indicate the 90 per cent and 99 per cent confidence intervals of the
$\chi^2$ distribution. The shaded region indicates the 99 per cent
confidence interval for the ratio $(\sigma_{\parallel,\rm
G2}/\sigma_{\parallel, \rm G1})$, determined above. The dot gives the
ratio we determined from the $\sim$1 mag luminosity difference. We see
that this ratio lies well below the 99 per cent confidence level,
which shows that the ratio expected from the mass model agrees with
that determined from the F--J relation. Also the ratio from the T--F
relation agrees well.

\subsubsection{Time delay}

The predicted time delay between components A and B is
$1.51^{+0.65}_{-0.60} \cdot f^{\rm b}_{\rm d,s}$ days (model II),
where $f^{\rm b}_{\rm d,s}=(1+z_{\rm d})\cdot [D_{\rm d}D_{\rm
s}/(D_{\rm ds}\;{\rm Gpc})]$ with $D_{\rm d}$, $D_{\rm ds}$ and
$D_{\rm s}$ being the angular diameter distances between
observer-lens, lens-source and observer-source, respectively and
$z_{\rm d}$ the redshift of the lens.  For typical lens (0.5) and
source (1.5) redshifts, the delay is around $7/h_{50}$ days (flat
universe with $\Omega_{\rm m}=1$). The velocity dispersion of G1 is
$97.6^{+5.4}_{-5.4}\cdot f^{\rm a}_{\rm d,s}$ km s$^{-1}$ (model II),
where $f^{\rm a}_{\rm d,s}=\sqrt{{D_{\rm s}}/{D_{\rm ds}}}$.  The
errors indicate the 99 per cent confidence interval, inferred from the
Monte-Carlo simulations. The 68 per cent (1$\sigma$) confidence
intervals are $\sim$2.5 times smaller.  Both the velocity dispersion
and time delay depend on the chosen cosmological model through the
angular diameter distances.  A good description of the dependence of
the model time delay and the angular diameter distances on the
cosmologcal model is given in Helbig (1997). In a flat universe
($\Omega_{\rm m}+\Omega_{\Lambda}=1$) the difference in time delay is
$\la$10 per cent between $\Omega_{\rm m}=1$ and $\Omega_{\rm
m}=0.3$. For a non-flat universe significant ($\gg$10 per cent)
differences in time delay are possible, depending on the clumpiness of
matter and the combination of $\Omega_{\Lambda}$ (normalised
cosmological constant) and $\Omega_{\rm m}$ (matter density).  All
other parameters in Table 4 are dimensionless and independent of the
cosmological model.

We conclude that model II is in substantial agreement with all
available radio and optical observations of B1127+385. However, more
detailed studies of the SB distributions of G1 and G2 are necessary to
improve the models and tighten the confidence intervals.  Moreover
both lens and source redshifts are unknown.

\begin{table}
  \caption{The minimum$-\chi^2$ mass model parameters for
    B1127+385. Listed are the SD axis ratio $(b/a)_{\Sigma}$, the velocity
    dispersion $\sigma_{\parallel}$, the position angle (PA) of the major
    axis and the centre $(x,y)_{\rm G1}$ of G1 and the position and
    velocity dispersion of G2. The errors on axis ratio, velocity
    dispersion and position angle indicates the 99 per cent confidence
    interval inferred from Monte-Carlo simulations (Fig. 6).  $(x,y)_{\rm
      src 1/2}$ give the source positions of the subcomponents A1\&B1 and
    A2\&B2, respectively. Furthermore, $r_{A/B}$ and $\mu_{\rm A, B}$ give
    the derived flux density ratios and image magnifications. The
    coordinate system has been centered on image A1. The definitions of
    $f^{\rm a}_{\rm d,s}$ and $f^{\rm b}_{\rm d,s}$ are given in Section
    4.4.2.}
  \centering
  \begin{tabular}{lll}
    & Model I & Model II \\ $(b/a)_{\Sigma}^{\rm G1}$ &
    0.56 & $0.74^{+0.10}_{-0.12}$ \\ $(x,y)_{\rm G1}$ (mas) & (-228.0, &
    (-228.0, \\ & -38.0) & -38.0) \\ $\sigma_{\parallel}^{\rm G1}$ (km s$^{-1}$) &
    $108.9\cdot f^{\rm a}_{\rm d,s}$ & $97.6^{+5.4}_{-5.4}\cdot f^{\rm a}_{\rm d,s}$
    \\ PA$_{\rm G1}$ ($^\circ$) & -26.4 & $4.9^{+28.2}_{-22.4}$ \\ 
    $(x,y)_{\rm G2}$ (mas) & --- & (-749.0,\\ & --- & -337.0)\\ 
    $\sigma^{\rm G2}_{\parallel}$ (km s$^{-1}$) & --- & $0.79\cdot\sigma_{\parallel}^{\rm
      G1}$ \\ $(x,y)_{\rm src 1}$ (mas) & -350.20$\pm$0.03, &
    -439.81$\pm$0.03,\\ & -1.10$\pm$0.07 & -113.94$\pm$0.02 \\ 
    $(x,y)_{\rm src 2}$ (mas) & -350.38$\pm$0.03, & -439.89$\pm$0.03, \\ 
    & -1.62$\pm$0.07 & -114.42$\pm$0.02 \\ $r_{\rm B1/A1}$ & 0.56
    & 0.71 \\ $r_{\rm B2/A2}$ & 0.57 & 0.71 \\ $\mu_{\rm A1,A2}$ & -3.73
    & -4.65 \\ & -3.66 & -4.60 \\ $\mu_{\rm B1,B2}$ & 2.09 & 3.28 \\ &
    2.09 & 3.28 \\ $\Delta t_{(\rm \tiny B1-A1)}$ (days) & $2.64
    \cdot f^{\rm b}_{\rm d,s}$ & $1.51^{+0.65}_{-0.60} \cdot f^{\rm b}_{\rm d,s}$ \\ 
    $\chi^2/{\rm NDF}$ & 4.5 & 1.1 \\
  \end{tabular}
\end{table}

\begin{figure}
\begin{center}
  \leavevmode
\vbox{%
  \epsfysize=5.0cm \epsffile{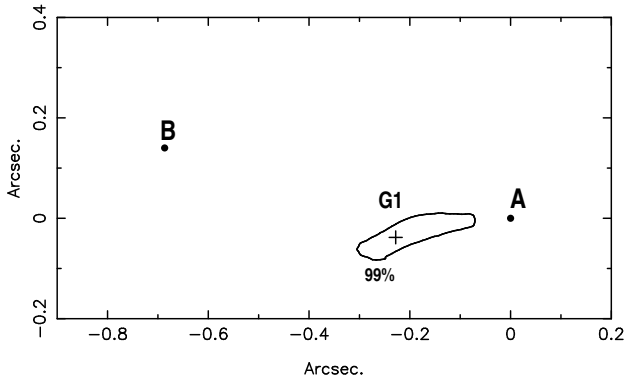}}
\end{center}
\caption{The 99 per cent confidence contour of the central SD position of G1,
  determined only on the basis of the VLBA radio images.
  The cross gives the 99 per cent confidence region of the SB distribution
  of G1 relative to the optical component west of it. The two circles
  indicate the two radio components, assuming they are associated with
  the optical emission features (see text). With this assumption, the
  SB centre of G1 is in excellent agreement with its inferred SD centre.}
\end{figure}

\begin{figure*}
\begin{center}
  \leavevmode
\vbox{%
  \epsfysize=15.0cm \epsffile{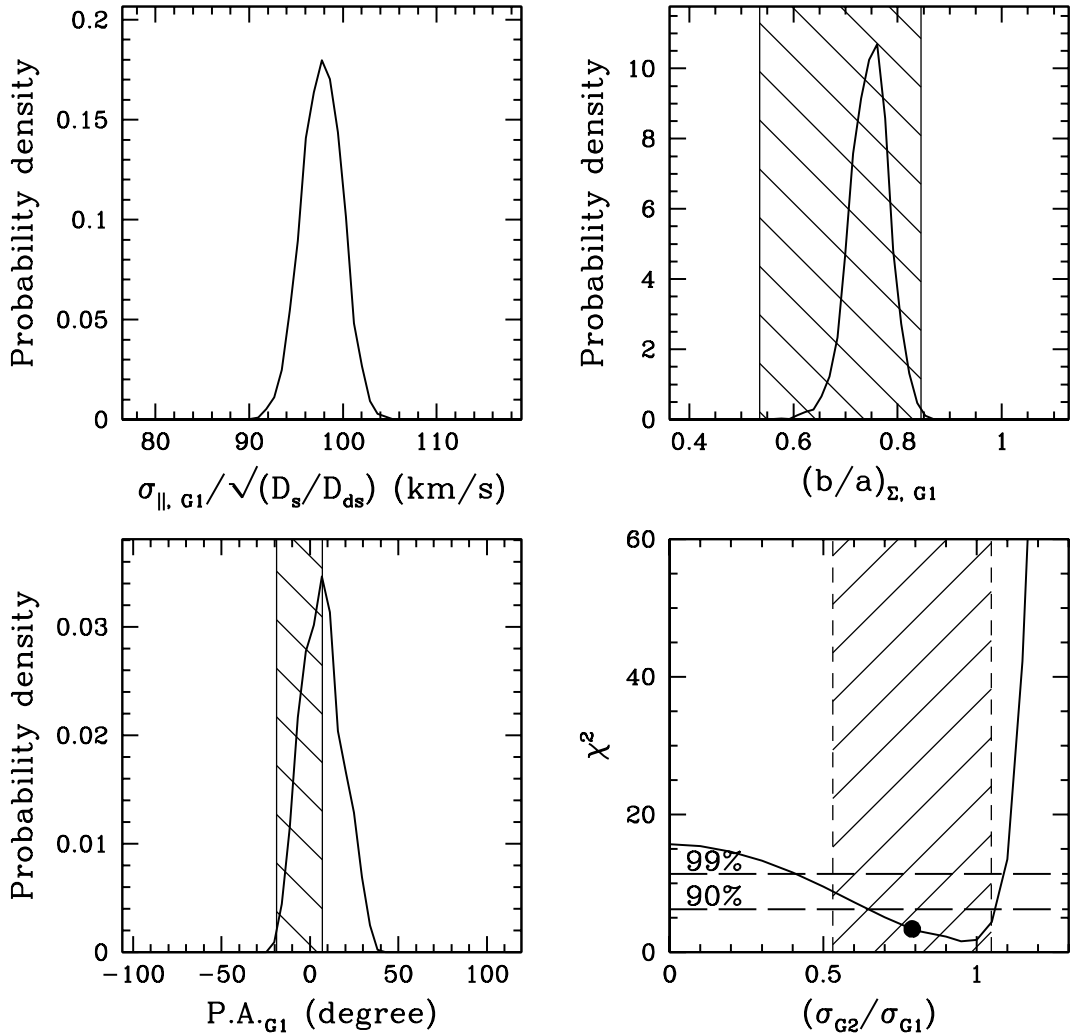}}
\end{center}
\caption{From upper left to lower right: Monte-Carlo probability density
  distributions (of models with $\chi^2<11.3$; 99 per cent confidence interval) of
  $\sigma_{\parallel, \rm G1}$, $(b/a)_{\Sigma, \rm G1}$ and P.A.$_{\rm G1}$.
  The shaded regions (99 per cent confidence interval) indicate these
  parameters as derived from the SB distribution of G1. The lower
  right panel shows $\chi^2$ as function of $(\sigma_{\parallel,\rm
    G2}/\sigma_{\parallel, \rm G1})$.  The observationally derived ratio
  falls well within the region containing 90 per cent of the $\chi^2$
  distribution.}
\end{figure*}

\subsection{Galaxy colours, luminosities and ${\rm M}/{\rm L}$ ratios} 

Having convinced ourselves that G1 and G2 are the lens galaxies, we
compare their colours (F555W--F814W$\approx$1.9--2.0) with the
synthesized galaxy colours in Fukugita, Shimasaku \& Ichikawa
(1995). For the different galaxy types approximate photometric
redshifts of 0.3 (E), 0.4 (S0), 0.4 (Sab), 0.7 (Sbc) and 0.9 (Scd) are
found.  G1 and G2 could therefore be early type galaxies (including
Sab) at low redshift (0.3--0.4) or late type galaxies at high redshift
(0.7--0.9). The integrated luminosities of G1 and G2 in $B$ band for
the different types of galaxies are: $\log_{10}(L_{{\rm B}\odot})
\approx 8.4-2\log(h_{50})$ (E), $8.8-2\log(h_{50})$ (S0 and Sab),
$9.6-2\log(h_{50})$ (Sbc) and $9.8-2\log(h_{50})$ (Scd) for G1 (${\rm
H}_{0}=50\cdot h_{50}$ km s$^{-1}$ Mpc$^{-1}$), where we used the
photometric redshifts found above and the $B$--F814W colours and K
corrections from Fukugita et al. (1995). For G2 these values are 0.4
lower.

At a redshift $z_{\rm d}=0.3$, $I$=22.5 (G1) corresponds to an
absolute magnitude $M_I \sim-19-5\log(h_{50})$. For an E and S0 type
galaxy this would mean it is $\sim$4 mag underluminous compared to E
and S0 type galaxies in the Hubble Deep Field (Mobasher et
al. 1996). Placing the galaxy at higher redshifts would make the $V-I$
colours of G1 and G2 inconsistent with those of E or S0 type galaxies
(Fukugita et al. 1995). The absolute $I$ magnitude is consistent
however with somewhat later type spiral galaxies at higher redshifts
($z_{\rm d}\ga0.5$). This would also explain why the velocity
dispersions of G1 and G2 appears significantly smaller than those
expected for $L_*$ E and S0 type galaxies (e.g. Kochanek 1993, 1994).

Using the velocity dispersions listed in Table 4 (model II), the
mass-to-light ratios (using the mass inside the Einstein radius, the
photometric lens redshifts and the total B luminosity) of G1 and G2
are: $(M/L_{{\rm B}}) \sim 60\cdot h_{50}\; {\rm M}_{\odot}/{\rm
L}_{\rm B, \odot}$ (E), $\sim 35\cdot h_{50}$ (S0 and Sab),
$\sim10\cdot h_{50}$ (Sbc) and $\sim 10\cdot h_{50}$ (Scd) (assuming
$z_{\rm s}=1.5$). All of these mass-to-light ratios are significantly
larger than normal galaxies of similar type, except for the higher
redshift late-type spiral galaxies. Dust obscuration could increase
the mass-to-light ratio, however it would also make the F555W--F814W
colours less reasonable.

We should note however that at intermediate redshift an error of 0.2
in redshift introduces $\sim$1 mag error in B and as mentioned, dust
obscuration (e.g. B1600+434; Jaunsen \& Hjorth 1997; Koopmans, de
Bruyn \& Jackson 1998) or luminosity evolution (e.g. Bender, Ziegler
\& Bruzual 1996; Hudson et al. 1998) have not been taken into account.
 
The results above therefore give only an indication. Redshifts of G1
and G2, and more accurate colours are vital to distinguish between
galaxy types and their mass-to-light ratios. To obtain the redshifts
of these galaxies in a reasonable integration time, one requires an 8
or 10-m class telescope (e.g. the Very Large Telescope (VLT) or Keck).

\subsection{Source}

We estimate ${\rm I}\sim$24.5 for the optical emission associated with
radio component A. Correcting for a magnification of 4.6 (Table 4),
this corresponds to an intrinsic I$\sim$26, a luminosity of
$\log_{10}(L_{{\rm I}\odot})\sim 9.8-2\log(h_{50})$ at $z_{\rm s}=1.5$
and an absolute magnitude $M_I \sim -19-5\log(h_{50})$ at $z_{\rm
s}=1.5$ (no K-corrections applied).

Comparing this with the absolute $I$ magnitudes of galaxies in the
Hubble Deep Field, the source is most consistent with a spiral galaxy
of type Sbc or later, as most E and S0 type galaxies have $M_ I
\la-23$ (Mobasher et al. 1996). At a redshift of $\sim$3 the
luminosity increases to $\log_{10}(L_{{\rm I}\odot})\sim
10.5-2\log(h_{50})$, still in the range of spiral galaxies (type Sab
or later; Mobasher et al. 1996).  We should note here that these
values have not been corrected for evolution or dust absorption,
internally or by the lens galaxies (G1 and G2).

However, the CLASS lens systems B0712+472 (Jackson et al. 1998) and
B1933+503 (Sykes et al. 1998; Jackson, private communication) also
appear to have very low luminosity sources, which could indicate that
a significant fraction of the weak (few mJy) flat spectrum radio
sources are associated with low luminosity objects, possibly late-type
spiral galaxies.

\section{Conclusions}

A new gravitational lens system with two images separated by $701\pm1$
mas has been discovered in the CLASS survey. The two radio components
have a flat spectrum between 0.327 GHz (WSRT; WENSS), 1.7 GHz
(MERLIN), 5.0 GHz (MERLIN) and 8.4 GHz (VLA). VLBA observations show
substructure in both images. An HST $I$-band image reveals two
emission features close to the radio components, which we identify
with two lens galaxies.  The colours and mass-to-light ratios of these
galaxies seem to favour two late-type spiral galaxies at relatively
high redshifts ($z_{\rm d}\ga0.5$).

The VLBA radio structure and optical HST emission are consistent with
a two-lens mass model, where a SIE and SIS mass distribution are
placed on the SB centres of G1 and G2, respectively.  This model is
able to reproduce the separations and position angles of the VLBA
substructure in radio components A and B, and their flux density
ratios. Our best model has a reduced $\chi^2$ of 1.1. Assuming both
lens galaxies are spiral galaxies, $\chi^2$ slightly increases to
1.6. Omitting G2 in the mass model increases the reduced $\chi^2$ to
4.5, which can therefore be excluded as an appropriate model with 99.6
per cent confidence.

Our best model gives a SD axis ratio of $0.74^{+0.10}_{-0.12}$,
position angle of $(4.9^{+28.2}_{-22.4})^\circ$ and velocity
dispersion of $97.6^{+5.4}_{-5.4}\cdot f^{\rm a}_{\rm d,s}$ km
s$^{-1}$.  The predicted time delay between radio components A and B
is $1.51^{+0.65}_{-0.60} \cdot f^{\rm b}_{\rm d,s}$ days.  For a
typical lens (0.5) and source redshift (1.5), a time-delay of $\sim$7
days is expected for H$_0$=50 km s$^{-1}$ Mpc$^{-1}$ and $\Omega_{\rm
m}=1$ in a flat universe.  WSRT monitoring data is in hand, to see if
the lensed object in B1127+385 is variable and therefore useful for
determining a time-delay between the lensed images. This time-delay
can constrain the Hubble parameter (Refsdal 1964).  The errors on
these parameters indicate the 99 per cent confidence interval of the
probability density distributions found from Monte-Carlo simulations.

Having constructed a consistent model within the lensing hypothesis --
explaining both the available radio and optical data of B1127+385 --
it appears that B1127+385 most likely is a new gravitational lens
system.

\section*{Acknowledgments}

LVEK and AGdeB acknowledge the support from an NWO
program subsidy (grant number 781-76-101). This research was supported
in part by the European Commission, TMR Programme, Research Network
Contract ERBFMRXCT96-0034 `CERES'. The National Radio Astronomy
Observatory is a facility of the National Science Foundation operated
under cooperative agreement by Associated Universities, Inc. MERLIN is
a national UK facility operated by the University of Manchester on behalf
of PPARC. This research used observations with the Hubble Space
Telescope, obtained at the Space Telescope Science Institute, which is
operated by Associated Universities for Research in Astronomy Inc.
under NASA contract NAS5-26555. The Westerbork Synthesis Radio
Telescope (WSRT) is operated by the Netherlands Foundation for
Research in Astronomy (ASTRON) with the financial support from the
Netherlands Organization for Scientific Research (NWO).


\begin{thebibliography}{}

\bibitem[\protect\citename{Bender, Ziegler \& Bruzual}1996]{BZB}
  Bender R., Ziegler B., Bruzual G., 1996, ApJL, 463, 51

\bibitem[\protect\citename{Faber \& Jackson}1976]{FJ} Faber S.M.,
  Jackson R.E., 1976, ApJ, 204, 668

\bibitem[\protect\citename{Fukugita \& Turner}1991]{FT} Fukugita M.,
  Turner E.L., 1991, MNRAS, 253, 99

\bibitem[\protect\citename{Fukugita, Shimasaku \& Ichiwaka}1995]{fuku}
  Fukugita M., Shimasaku K., Ichiwaka T., 1995, PASP, 107, 945

\bibitem[\protect\citename{Helbig}1997]{helbig}
  Helbig, P., Proceedings of the Workshop on Golden Lenses, 1997,
  http://multivac.jb.man.ac.uk:8000/ceres/\\
  workshop1/proceedings.html

\bibitem[\protect\citename{Hudson et al.}1998]{Hudson} Hudson M.J.,
  Gwyn S.D.J., Dahle H., Kaiser N., ApJ, 503, 531

\bibitem[\protect\citename{Jackson}1998]{jackson} Jackson N., et al.,
  1998, MNRAS, 296, 483

\bibitem[\protect\citename{Jaunsen \& Hjorth}1996]{jauns} Jaunsen
  A.O., Hjorth J., 1997, A\&A, 317, L39

\bibitem[\protect\citename{Kayser}1990]{kays1} Kayser R., 1990, ApJ,
  357, 309

\bibitem[\protect\citename{Kochanek}1991]{koch1} Kochanek C.S., 1991,
  ApJ, 379, 517

\bibitem[\protect\citename{Kochanek}1993]{koch2} Kochanek C.S., 1993,
  ApJ, 419, 12

\bibitem[\protect\citename{Kochanek}1994]{koch3} Kochanek C.S., 1994,
  ApJ, 436, 56

\bibitem[\protect\citename{Kochanek}1996]{koch4} Kochanek C.S., 1996,
  ApJ, 466, 638

\bibitem[\protect\citename{Koopmans}1998]{koopm} Koopmans L.V.E., de
  Bruyn, A.G., Jackson N., 1998, MNRAS, 295, 534

\bibitem[\protect\citename{Kormann, Schneider \&
    Bartelmann}1994]{korr} Kormann R., Schneider P., Bartelmann M.,
  1994, A\&A, 284, 285

\bibitem[\protect\citename{Mobasher}1996]{mobasher} Mobasher B., 
  Rowan-Robinson M., Georgakakis A., Eaton N., 1996, MNRAS, 282, 7L

\bibitem[\protect\citename{Patnaik}1992]{patnaik} Patnaik A.R., 
  Browne I.W.A., Wilkinson P.N., Wrobel J.M., 1992, MNRAS, 254, 655

\bibitem[\protect\citename{Pearson}1994]{pearson} Pearson T.J., 
  Shepherd M.C., Taylor G.B., Meyers S.T., 1994, BAAS, 185, $\#$08.08

\bibitem[\protect\citename{Refsdal}1964]{Refs} Refsdal S., 1964,
  MNRAS, 128, 295

\bibitem[\protect\citename{Refsdal}1964]{Refs} Refsdal S., 1966,
  MNRAS, 134, 315

\bibitem[\protect\citename{Rhee}1996]{rhee} Rhee M-H., 1996, PhD
  Thesis, University of Groningen

\bibitem[\protect\citename{Schneider, Ehlers \& Falco }1992]{schn}
  Schneider P., Ehlers J., Falco E.E., 1992, Gravitational Lenses,
  Springer Verlag, Berlin

\bibitem[\protect\citename{Shepherd}1997]{shep} Shepherd M.C., 1997,
  ADASS VI, A.S.P Conference Series, vol 125, eds., Gareth Hunt and
  H.E. Payne, p77

\bibitem[\protect\citename{Sykes}1997]{sykes} Sykes C.M., et al.,
  1997, MNRAS, astro-ph/9710358

\bibitem[\protect\citename{Thompson}1986]{thompson} Thompson A.R.,
  Moran J.M., Swenson G.W., 1986, Interferometry and Synthesis in 
  Radio Astronomy, Wiley-Interscience publication, p262

\bibitem[\protect\citename{Tully}1977]{tully} Tully R.B., 
  Fisher J.R., 1977, A\&A, 54, 661

\end{thebibliography}
\end{document}